\documentclass[twocolumn,showpacs,preprintnumbers,amsmath,amssymb,10pt,fleqn]{revtex4}

\usepackage{graphicx}
\usepackage{dcolumn}
\usepackage{bm}
\usepackage{hyperref}

\begin{document}
\title{Internal structure of a Maxwell-Gauss-Bonnet black hole}

\author{S. Alexeyev}
 \email{alexeyev@sai.msu.ru}
 \affiliation{%
Sternberg Astronomical Institute, Lomonosov Moscow State
University, Universitetsky Prospekt, 13, Moscow 119991, Russia}%
\author{A. Barrau}%
 \email{aurelien.barrau@cern.ch}
\affiliation{%
Laboratoire de Physique Subatomique et de Cosmologie, UJF-INPG-CNRS\\
53, avenue des Martyrs, 38026 Grenoble cedex, France\\
{\rm and}\\
Institut des Hautes Etudes Scientifiques\\
35, route de Chartres, 91440, Bures-sur-Yvette}%
\author{K.A. Rannu}%
 \email{melruin1986@gmail.com}
\affiliation{%
Sternberg Astronomical Institute, Lomonosov Moscow State
University, Universitetsky Prospekt, 13,Moscow 119991, Russia}%
\date{\today}

\begin{abstract}
The influence of the Maxwell field on a static, asymptotically flat
and spherically-symmetric Gauss-Bonnet black hole is considered.
Numerical computations suggest that if the charge increases beyond a
critical value, the inner determinant singularity is replaced by an
inner singular horizon.
\end{abstract}

\maketitle

The study of new physical effects induced by the 4-dimensional
low-energy effective string action with second order curvature
correction has been an important topic in black hole physics during
the last three decades (see, {\it e.g.}, \cite{action}). The internal
structure of black holes described by the action
\begin{eqnarray}
S &=& \frac{1}{16\pi} \int d^4 x \sqrt{-g} \left[ m^2_{pl} (- R +
2\partial_{\mu} \phi \partial^\mu \phi)\right. \nonumber \\
&&\left.- e^{-2\phi}
F_{\mu\nu} F^{\mu\nu} +
\lambda e^{-2\phi} S_{GB} \right], \label{eq:01}
\end{eqnarray}
where $m_{pl}$ is the Plank mass, $\phi$ is the dilaton field, $R$ is
the scalar curvature, $S_{GB} = R_{ijkl} R^{ijkl} - 4 R_{ij}R^{ij} +
R^2$ is the Gauss-Bonnet term, $F_{\mu\nu}F^{\mu\nu}$ is the Maxwell
field and $\lambda$ is the string coupling constant, has been
investigated in \cite{alex1}. The influence of the magnetic charge of
the black hole on the behavior of the metric functions was considered
and it was shown that there exists a "critical value" of the charge
beyond which the influence of the Maxwell term becomes more important
than the Gauss-Bonnet one. The inner determinant singularity at
$r=r_s$ is then replaced by a smooth local minimum.

In this paper, we focus on the behavior of the curvature invariant
$R_{ijkl}R^{ijkl}$ near this critical point and in the vicinity of the
main singularity at $r=r_x$.

Considering a static, asymptotically flat and spherically symmetric
black hole solution, we focus on the following metric:
\begin{equation}
{ds^2} = {\Delta dt^2} - \frac{\sigma^2}{\Delta} dr^2 -
f^2 (d\theta^2 + {\sin}^2{\theta}d {\varphi}^2),
\label{eq:02}
\end{equation}
where $\Delta$, $\sigma$ and $f$ are functions that depend on the
radial coordinate $r$ only. To simplify the problem, only the magnetic
charge will be taken into account. Therefore, for the Maxwell tensor
$F_{\mu\nu}$, one can use the ansatz $F = q \sin\theta \ d\theta \wedge
d\varphi$ \cite{ghs}. The corresponding field equations in the GHS
gauge ($\sigma(r)=1$) are as follows:
\begin{widetext}
\begin{eqnarray}
& m_{Pl}^2 & [f f'' + f^2 (\phi')^2 ] + 4 e^{-2\phi} \lambda
[\phi'' - 2(\phi')^2]
{\Delta (f')^2 - 1} + 4 e^{-2\phi}
\lambda \phi' 2 \Delta f' f'' = 0,
\label{eq:03} \\
& m_{Pl}^2 & [1 + \Delta f^2 (\phi')^2 - \Delta' f f'
- \Delta (f')^2 ] +
4 e^{-2\phi} \lambda \Delta' \phi' [1 -
3 \Delta (f')^2 ] - \nonumber \\
&& - {\ } e^{-2\phi} q^2 f^{-2} = 0, \label{eq:04} \\
& m_{Pl}^2 & [\Delta'' f + 2 \Delta' f'
+ 2 \Delta f'' + 2 \Delta f (\phi')^2] +
4 e^{-2\phi} \lambda[\phi'' - 2 (\phi')^2]
2 \Delta \Delta' f' + \nonumber \\
&& + {\ } 4 e^{-2\phi} \lambda \phi' 2
[(\Delta')^2 f' + \Delta \Delta'' f' +
\Delta \Delta' f''] - 2 e^{-2\phi} q^2 f^{-3}
= 0, \label{eq:05} \\
- 2 & m_{Pl}^2 & [\Delta' f^2 \phi' + 2 \Delta f f'
\phi' + \Delta f^2 \phi''] +
4 e^{-2\phi} \lambda [(\Delta')^2 (f')^2
+ \Delta \Delta'' (f')^2 + \nonumber \\
&& + {\ } 2 \Delta \Delta' f' f''- \Delta'']
- 2 e^{-2\phi} q^2 f^{-2}= 0.
\label{eq:06}
\end{eqnarray}
\end{widetext}

The behavior of the metric functions and of the dilatonic field near
the horizon are described by a simple Taylor expansion
\cite{alex2}:
\begin{eqnarray}\label{eq:07}
\Delta & = & d_1 x + d_2 x^2 + O(x^2), \nonumber \\
f & = & f_0 + f_1 x + f_2 x^2 + O(x^2), \\
e^{-2\phi} & = & e^{-2\phi_0} + \phi_1 x + \phi_2 x^2 + O(x^2), \nonumber
\end{eqnarray}
where ($x = r - r_h, \ll 1$).\\

Without the Gauss-Bonnet term, the
Gibbons-Maeda-Garfinkle-Horowitz-Strominger solution (GM-GHS)
\cite{ghs} should be recovered as the basic solution of the Einstein
equations with the dilaton and Maxwell terms. This solution is given
by:
\begin{eqnarray}
d s^2 & = & \left( 1 - \frac{2M}{r} \right) d t^2
- \left( 1 - \frac{2M}{r} \right)^{-1} d r^2 \nonumber \\
&&- r \left( r - \frac{q^2 \exp (2 \phi_0)}{M} \right)
d \Omega, \nonumber \\
&& \exp( -2 \phi) = \exp ( -2 \phi_0 ) - \frac{q^2}{Mr},
\label{eq:08}
\end{eqnarray}
where $M$ stands for the black hole mass. In the limit $\lambda \to 0$,
the solution of equations (\ref{eq:03})--(\ref{eq:06}) at infinity
should coincide with Eq.~(\ref{eq:08}).

In order to determine the two metric functions and the dilatonic
field, three equations are required. Among the four equations
(\ref{eq:03})-(\ref{eq:06}), only equations (\ref{eq:03}),
(\ref{eq:05}) and (\ref{eq:06}), which contain the second derivative
of the metric functions and the dilaton, are used. In contrast,
Eq.~(\ref{eq:04}), which contains the first derivative only, is considered
as a constraint to check the solution.

To solve the system (\ref{eq:03})-(\ref{eq:05})-(\ref{eq:06}), the
equations are rewritten using $E=e^{-2 \phi}$ instead of the dilaton
itself. Furthermore, the case $\lambda=1$ is considered. In the chosen
metric gauge, the squared Riemann tensor is given by:
\begin{eqnarray}\label{eq:09}
R_{ijkl}R^{ijkl} &=& {\Delta''}^2
+ 4 {\Delta'}^2 \frac{{f'}^2}{f^2}
+ 8 \Delta^2 \frac{{f''}^2}{f^2}
+ 8 \Delta \Delta' \frac{f' f''}{f^2} \nonumber\\
&+& \frac 4{f^4} -
8 \Delta \frac{{f'}^2}{f^4}
+ 4 \Delta^2 \frac{{f'}^4}{f^4}.
\end{eqnarray}

The main difficulty in solving the system numerically is the fact that
the metric function $\Delta$ has a coordinate singularity at the event
horizon, making the numerical calculation "through" the horizon
intricate. This is why the computation process was divided into two
parts. First, the GM-GHS solution (\ref{eq:08}) was taken as the
initial condition at infinity. Solutions for the metric functions and
the dilaton outside the event horizon were found. Then, the results
near the horizon were taken as new initial conditions.

\begin{figure}[ht]
\begin{center}
\includegraphics[scale=0.35]{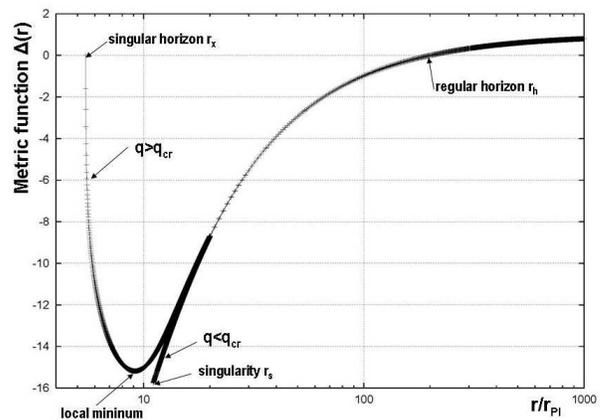}
\end{center}
\caption{Metric function $\Delta$ as a function of the radial coordinate
$r$ for $q=21.50<q_{cr}$ (left curve) and
$q=24.81>q_{cr}$ (right curve) when $r_h=200.0$ Planck units.}
\label{f:01}
\end{figure}

The behavior of metric the functions $\Delta$ and $f$, together with
the dilaton exponent $e^{-2\phi}$, were investigated under the event
horizon of the black hole. It differs significantly depending on the
black hole charge. If the charge is zero or small the metric function
$\Delta$ is defined only for $r>r_s$ ($r_s$ being smaller than the
event horizon radius $r_h$). In this case, there exist two
mathematical branches: one is physical (and displayed on
Fig.~\ref{f:01}), ranging from $r=r_s$ to infinity, and the other one
is an artifact, ranging from $r=r_s$ to $r=r_x$. If the value of the
charge is larger than a critical value $q_{cr}$, the inner singularity
does not exist anymore and, as it can be seen on Fig.~\ref{f:01}
(right), $\Delta$ exhibits a local minimum. When the black hole charge
increases from zero, a phase transition occurs at $q=q_{cr}$ such that
the inner singularity disappears (being relpaced by a local minimum
for $\Delta$) and an inner horizon forms at $r_x$. This is the main
difference between the considered solution and the GM-GHS case.

The values of the critical charge have been numerically computed for
different masses and are given in Table~\ref{t:3}.
\begin{table}[hb]
  \begin{center}
    \begin{tabular}[c]{|c|c|c|c|c|c|c|c|c|c|}
      \hline
      & & & & & & & &\\
      $M$ & $5.0$ & $10.0$ & $20.0$ & $50.0$ & $60.0$ & $70.0$ & $90.0$ & $100.0$\\
      & & & & & & & &\\
      \hline
      & & & & & & & &\\
      $q_{cr}$ & $4.53$ & $7.20$ & $11.42$ & $21.03$ & $23.75$ & $26.32$ & $31.13$ & $33.39$\\
      & & & & & & & &\\
      \hline
    \end{tabular}
  \end{center}
\caption{Black hole critical charge $q_{cr}$ as a function of the
mass $M$}\label{t:3}
\end{table}

\begin{widetext}

\begin{figure}[ht]
\begin{center}
\includegraphics[scale=0.35]{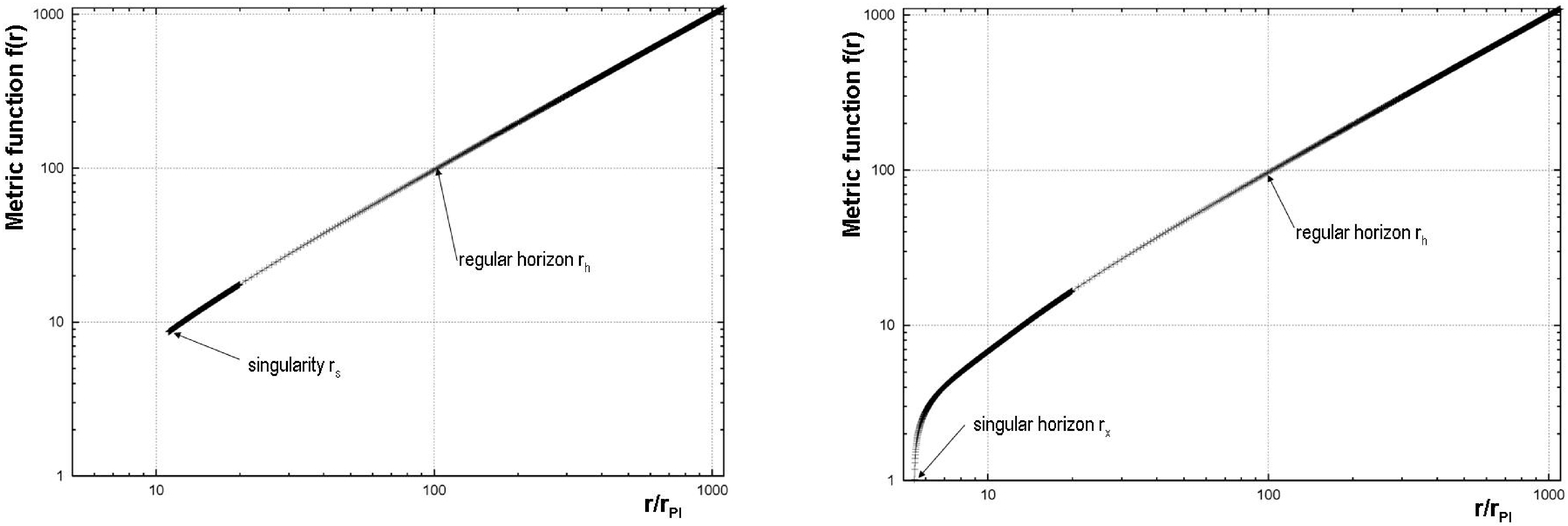}
\end{center}
\caption{Metric function $f$ as a function of the radial coordinate
$r$ for $q=21.50<q_{cr}$ (left plot) and $q=24.81>q_{cr}$ (right plot)
when $r_h=200.0$ Planck units.}
\label{f:02}
\end{figure}

\end{widetext}

\begin{figure}[ht]
\begin{center}
\includegraphics[scale=0.35]{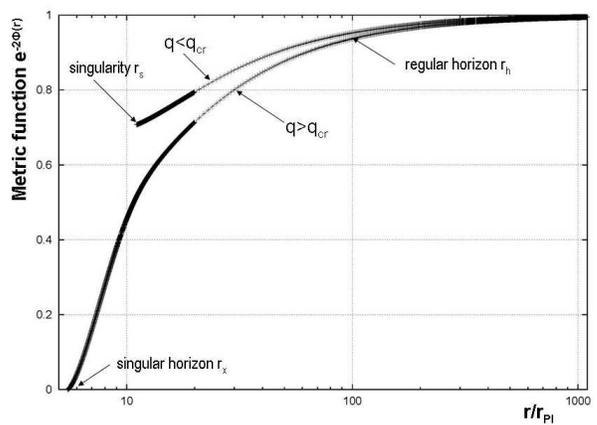}
\end{center}
\caption{Dilatonic exponent $e^{-2\phi}$ as a function of the radial
coordinate $r$ for $q=21.50<q_{cr}$ (upper curve) and $q=24.81>q_{cr}$
(lower curve) when $r_h=200.0$ Planck units.}
\label{f:03}
\end{figure}

The behavior of the metric function $f(r)$ (Fig.~\ref{f:02}) and
$e^{-2\phi(r)}$ (Fig.~\ref{f:03}) outside the horizon are analogous to
the GM-GHS case. For large values of the radial coordinate, $f(r) \sim
r$. When $q < q_{cr}$ these functions are monotonic from $r = r_s$ to
infinity. However, when $q > q_{cr}$, they are defined in a wider
interval $[r_x, \infty]$. The metric function $f(r)$ vanishes for
$r=r_x$, together with $e^{-2\phi}$. This underlines that for $r \to
r_x$, the influence of the Maxwell term becomes subdominant when
compared to the Gauss-Bonnet one.

\begin{widetext}

\begin{figure}[ht]
\begin{center}
\includegraphics[scale=0.45]{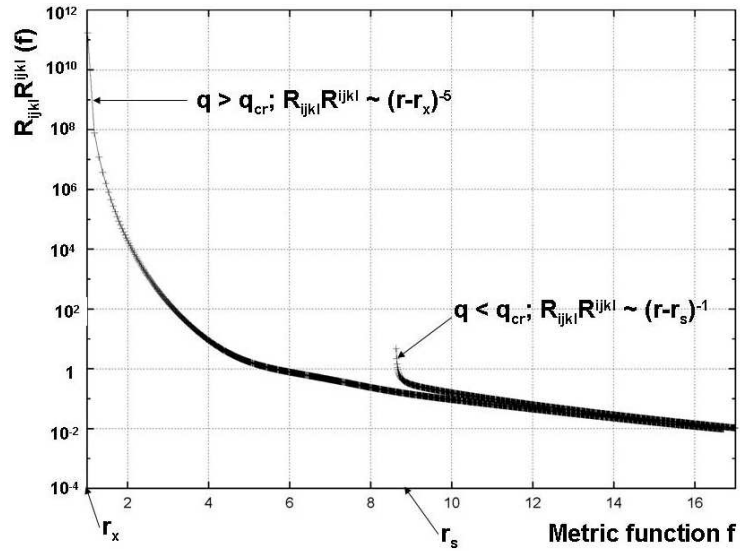}
\end{center}
\caption{Curvature invariant $R_{ijkl}R^{ijkl}$ as a function of the
metric function $f$ for $q=21.50<q_{cr}$ (left curve) and
$q=24.81>q_{cr}$ (right curve), with $r_h=200.0$ Planck units.}
\label{f:04}
\end{figure}


\begin{figure}[ht]
\begin{center}
\includegraphics[scale=0.35]{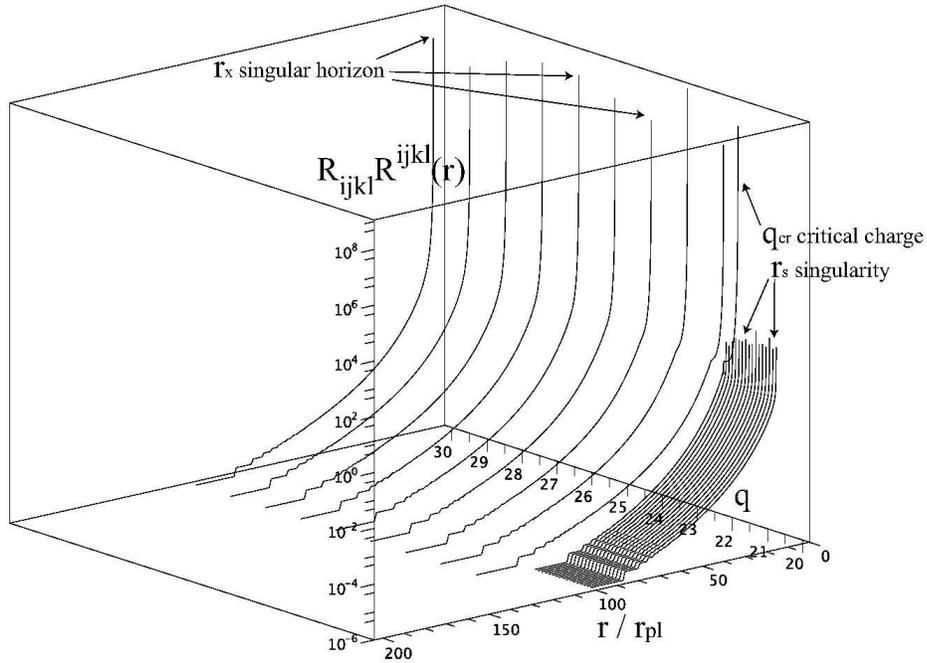}
\end{center}
\caption{Three dimensional dependence of the curvature invariant
$R_{ijkl}R^{ijkl}$ against the charge $q$ and the radial coordinate
$r$ for $r_h=200.0$.}
\label{f:05}
\end{figure}

\end{widetext}

The behavior of the curvature invariant $R_{ijkl}R^{ijkl}$ under the
event horizon of the black hole was also studied and it was confirmed
that $R_{ijkl}R^{ijkl} \to \infty$ for $r \to r_s$ when $q<q_{cr}$.
The situation when the black hole charge reaches its critical value
and the metric function $\Delta$ begins to exhibit a local minimum
instead of a singularity at $r = r_s$ was considered in more details.
It was checked that in this case the value of the curvature invariant
does not diverge anymore. It is therefore obvious that the local
minimum of the metric function $\Delta (r)$ is intrinsically non
singular.

When $q > q_{cr}$ ({\it i.e.} when the $r_s$ singularity vanishes),
the important point is $r_x$, where $f$ vanishes. It was numerically
checked that the curvature invariant diverges at this point. So, it
can be conjectured that $r = r_x$ becomes a singular horizon inside
the black hole. When $q < q_{cr}$, this horizon belongs to the
nonphysical branch of the considered system of equations. Near the
singular horizon $r_x$ (when $q>q_{cr}$), the curvature invariant
diverges significantly more rapidly than near the singularity $r_s$
(when $q < q_{cr}$).

In Fig.~\ref{f:04}, the curvature invariant is shown as a function the
the metric function $f$ which, in the chosen metric, has the intuitive
meaning of the radius of a two-sphere. In the same plot the asymptotic
dependence of the curvature invariant from the radial coordinate $r$
in the neibourhood of the discussed particular points is written down.
The asymptotic behavior of the metric function $f$ in those regions
can be expressed as follows \cite{alex1}:
\begin{eqnarray}\label{eq:10}
f(r \to r_s) & = & f_s + f_{s2} (r - r_s) + f_{s3} (\sqrt{r - r_s})^3
+ \ldots, \nonumber\\
f(r \to r_x) & = & f_x + f_{x1} \sqrt{r - r_x}
+ f_{x2} (r - r_x) + \ldots,
\end{eqnarray}
where $f_i$ are the numerical expansion coefficients.

So the divergence of the curvature invariant in terms of metric function
$f$ is given by
\begin{eqnarray}\label{eq:11}
R_{ijkl} R^{ijkl} \sim \mbox{const}_1 \times \left(f-f_s
\right)^{-1} \quad \mbox{for} \quad f \to f_s, \nonumber\\
R_{ijkl} R^{ijkl} \sim \mbox{const}_2 \times \left(f-f_x \right)^{-5}
\quad \mbox{for} \quad f \to f_x.
\end{eqnarray}

In Fig.~\ref{f:05}, the three-dimensional dependance of the curvature
invariant as a function of the charge $q$ and the radial coordinate $r$ 
are displayed.\\

This establishes the internal structure of a Maxwell-Gauss-Bonnet
black holes. It can also be noticed that the regularization of the
internal structure, which is expected by some models of ``cosmological
natural selection'' \cite{smolin} and is predicted by loop quantum
gravity \cite{bojo}, does not happen in Gauss-Bonnet gravity, even for
highly charged black holes.

\section*{Acknowledgments}

S.A. would like to thank Russian Foundation for Basic Research for the
support via grant No. 07-02-01034-a.

\end{document}